\documentclass{aa}
\usepackage{graphicx}
\newcommand {\beq}{\begin{equation}}
\newcommand {\eeq}{\end{equation}}
\newcommand {\civ}{C\,{\sc iv}}

\begin{document}

   \title{The non-constant slope of the \civ~Baldwin effect in  NGC 4151}

   \author{Minzhi Kong
   \inst{1,2,3}
   \and
   Xue-Bing Wu
   \inst{2}
  \and   Ran Wang
   \inst{2}
  \and
   Fukun Liu
   \inst{2}
   \and
   J.L. Han
   \inst{1}}

\offprints{XBW, \email{wuxb@bac.pku.edu.cn}}

   \institute{National Astronomical Observatories, Chinese Academy of Sciences, Beijing 100012, China
   \and
Department of Astronomy, Peking University, Beijing 100871, China
   \and
   Department of Physics, Heibei Normal University, Shijiazhuang
   050016, China
              }
\date{Received ; accepted }
\authorrunning{Kong et al.}
\titlerunning{ The \civ~Baldwin effect in NGC 4151}

\abstract {The relationship between the emission-line equivalent
width and the continuum luminosity, called the Baldwin effect, is
important for studies of the physics of the broad-line region of
AGNs. Some recent studies have revealed the non-constant slope of
the intrinsic Baldwin effect for several Seyfert 1 galaxies.}
{Using the archived ultraviolet spectra obtained by IUE, HST, and
HUT in  1978--2002, we investigated the intrinsic \civ~Baldwin
effect of the well-studied Seyfert 1 galaxy NGC 4151. Both its
continuum flux and \civ~emission-line flux varied about two orders
of magnitude in more than two decades, making it one of the best
targets for studying the slope variations caused by the Baldwin
effect.}
{We fitted the \civ~line profile of the 490 archived UV spectra of NGC 4151
 with a similar
model consisting of a few Gaussian components and derived the slope
in the log-log plot for the total flux of \civ~emission line against
the UV continuum flux in different observation epochs.}
{We found that the slope is not constant for NGC 4151, as it varies
from 0.58 in the highest flux epoch to 0.83 in the lowest flux
epoch. The slope evidently decreases as the continuum flux
increases, which reinforces the previous findings of the
non-constant slope in the H$_\beta$ Baldwin effect of NGC 5548 and
the \civ~Baldwin effect of Fairall 9. }
{Our result suggests that such a non-constant slope may not be
unusual for AGNs. Its physical origin is probably related to the
different non-linear responses of the emission line to the variable
ionizing continuum caused by the different accretion modes at
different luminosity levels. We briefly discuss the effects of
various absorption components in the \civ~line profile of NGC 4151
and argue that the slope variation is not driven mainly by the
absorption effect. Intensive, long-term, and high-resolution
spectral observations of strongly variable AGNs are needed in the
future to confirm our result.}


\maketitle
\section{Introduction}

The relation between the continuum and emission-line luminosities is
important for our understanding of AGN physics. Such a relationship,
known as the Baldwin effect, states that the broad emission-line
equivalent width (EW) decreases as the continuum luminosity
increases (Baldwin 1977; for a review see Osmer \& Shields 1999).
The Baldwin effect was first discovered for the \civ~$\lambda$1549
line (Baldwin 1977) and was confirmed later for most of the other
strong UV emission lines (Kinney, Rivolo \& Koratkar 1990), as well
for the optical hydrogen Balmer lines (Gilbert \& Peterson 2003;
Goad, Korista \& Knigge 2004).

The physical origin of the Baldwin effect is still unclear, but is
probably related to the luminosity-dependent continuum variations.
According to a theoretical work by Korista, Baldwin \& Ferland
(1998), the ionizing continuum shape is softened if the luminosity
increases. The reduction in the ionizing photons at a given
UV/optical luminosity leads to smaller equivalent widths of the
broad emission lines. Indeed, some accretion disk models can also
explain the continuum softening naturally in the case of increasing
luminosity (Netzer, Laor \& Gondhalekar 1992).  However, other
factors, such as the selection effect, variability, and light
travel-time effects, may also account for some parts of or the
scatters of the Baldwin effects (Jones \& Jones 1980; Murdoch 1983;
Krolik et al. 1991; Pogge \& Peterson 1992; Peterson et al. 2002).
Recent studies, including Baskin \& Laor (2004) and Bachev et al.
(2004), have revealed the possible correlations between the \civ~EW
and some parameters defining the Eigenvector 1 of AGNs (Boroson \&
Green 1992). Shang et al. (2003) also show that the Eigenvector 1
can contribute to the scatters of the Baldwin effect.

The Baldwin effect can be expressed by the simple formula $ EW
\propto L_c^{\beta} $, where  $L_c$ is the continuum luminosity, or
alternatively by $ L_{\rm line} \propto L_c^{\alpha}$, where $L_{\rm
line}$ is the emission line luminosity (evidently $\alpha=1+\beta$).
There are two kinds of Baldwin effects treated in the literature.
One is the {\it global} Baldwin effect, which is obtained from the
single-epoch observations for an ensemble of AGNs. The other is the
{\it intrinsic} Baldwin effect, which represents the line-continuum
relation in a single variable AGN. For the global Baldwin effect,
the $\beta$ value  was found to be $-0.17$ and $-0.12$ for \civ~and
Ly $\alpha$ lines, respectively (Kinney et al. 1990; Pogge \&
Peterson 1992). Some studies also indicate steeper slopes of the
global Baldwin effect for the relatively high ionization lines than
those found for the low ionization ones (Wu et al. 1983; Kinney et
al. 1987, 1990; Baldwin et al. 1989; Zheng et al. 1997). The
flattening of the slope of the global Baldwin effect at lower
continuum luminosities was mentioned previously by Osmer \& Shields
(1999) as a second-order effect, and was also noticed recently by
Baskin \& Laor (2004) from a study of 81 BQS quasars.

The intrinsic Baldwin effect (usually expressed as $ f_{line}
\propto f_c^{\alpha}$, where $f_{line}$ and $f_c$ are the emission
line and continuum fluxes, respectively), has been studied only for
several strongly variable AGNs. The derived $\alpha$ value varies
from 0.1 to 0.6 for different AGNs, with a mean value around 0.4
(Kinney et al. 1990; Pogge \& Peterson 1992). Recently, Gilbert \&
Peterson (2003) and Goad et al. (2004) studied the intrinsic Baldwin
effect for the best-studied Seyfert 1 galaxy NGC 5548 using the data
of 13-year-observations and found that the $\alpha$ value for the
H$\beta$ line varies in a range from 0.4 to 1.  Goad et al. (2004)
find that the $\alpha$ value decreases as the continuum flux
increases, suggesting the slope of the intrinsic Baldwin effect is
not constant. Wamsteker \& Colina (1986) and Osmer \& Shields (1999)
also notice the slope change of the intrinsic \civ~Baldwin effect
for another Seyfert 1 galaxy, Fairall 9, which clearly shows the
flattening of the EW value as the continuum flux decreases.
Interestingly, the trend of the slope change in the intrinsic
Baldwin effect seems consistent with that of the global Baldwin
effect. However, this may not indicate that the physics behind them
is totally the same. For example, either the different metallicity
or black-hole mass has been invoked as possible origin of at least
part of the global Baldwin effect (Warner, Hamann \& Dietrich 2004),
but this is clearly not relevant to the intrinsic Baldwin effect of
a single AGN where these parameters are fixed.

To further test the non-constant slope of the intrinsic Baldwin
effect, in this paper we investigate the relationship between the
\civ~emission line flux and the UV continuum flux for another
well-known nearby Seyfert 1 galaxy NGC 4151 that has been
extensively observed in the UV band by the {\it International
Ultraviolet Explorer} (IUE), {\it Hubble Space Telescope}
 (HST), and {\it Hopkins Ultraviolet Telescope} (HUT) in the past three decades
(Boksenberg et al. 1978; Clavel et al. 1987, 1990; Ulrich et al.
1991; Crenshaw et al. 1996, 2000; Kriss et al. 1992, 1995; Weymann
et al. 1997; Kraemer et al. 2001; see Ulrich 2000 for a detailed
review). Both the UV emission-line flux and continuum flux of NGC
4151 varied about two orders of magnitude in this long observation
period, making it one of the best targets for studying the variation
in the slope of the Baldwin effect.

In Sect.~2 we present the data analysis of the UV spectra from IUE,
HST, \& HUT. In Sect.~3 we show our result for the varying slope of
the intrinsic \civ~Baldwin effect. Finally we give our conclusions
and briefly discuss our result in Sect.~4.

\section{Data analysis of the archived UV spectra}
To investigate the slope variation of the intrinsic \civ~Baldwin
effect in NGC~4151, we measured the flux of the \civ~emission-line
and the flux of UV continuum observed in a long period.

\subsection{The data set}
As a nearby ($\rm cz=995\,km~s^{-1}$) bright Seyfert 1 galaxy, NGC
4151 has been extensively observed by IUE (Boksenberg et al. 1978;
Clavel et al. 1987, 1990; Ulrich 1996; Ulrich et al. 1991; Crenshaw
et al. 1996; Edelson et al. 1996). However, different results have
been found for the correlation between the UV continuum and the
emission-line flux variations. For example, from the observations in
the 1993 IUE campaign Crenshaw et al. (1996) found that the
emission-line light curve does not correlate well with the continuum
variation over the short duration of observations. But from the
long-term light curves over two decades, the response of the
emission-line flux with the continuum variation was clearly observed
(Ulrich et al. 1991; Ulrich 2000).  In this paper, we re-investigate
this correlation using the 468 archived low-resolution
large-aperture spectra of IUE/SWP from 1978 to 1996 (JD 2443722.59
to JD 2450243.40) where the \civ~emission line was clearly detected.

Furthermore, NGC~4151 was observed by HUT at a resolution of
3\,{\AA} in December 1990 and March 1995 (Kriss et al. 1992, 1995),
and the 8 archived HUT spectra are available. These observation were
done in a time period that partly overlapped with that of the IUE
observations.  Therefore, the HUT spectra can be used to test the
measurement accuracy of the IUE data.

NGC 4151 has also been frequently observed by HST. The 15 HST/STIS
archived spectra in 1998--2002 are available and used in our study.
However, the archived HST/GHRS spectra were skipped because they do
not cover the whole red part of the \civ~line profile (Weymann et
al. 1997).  In all of the 15 HST/STIS spectra, the whole \civ~line
profile is shown clearly.

In total, we collected 490 archived UV spectra of NGC 4151 with a
wavelength range from about 1300\,{\AA} to 1800\,{\AA}  observed by
IUE, HUT, \& HST in the period of 1978--2002.

\subsection{The continuum and \civ~emission-line flux determinations}

Low-resolution spectra of NGC 4151 taken with IUE and HUT clearly
show broad emission and remarkable central absorption in the
\civ~line profile. However, high-resolution spectra from HST have
revealed many more other subtle emission and absorption features
(Weymann et al. 1997; Crenshaw et al. 2000; Kraemer et al. 2001).
Therefore, the \civ~line profile of NGC 4151 is very complex and
needs to be treated with caution. Because 468 of the 490 archived
spectra in our study were obtained by IUE, the lower resolution of
IUE spectra does not allow a more accurate fitting of the subtle
emission and absorption features. In order to avoid the systematic
uncertainties caused by different fitting models, in this study we
simply adopt the same spectral-fitting model for the HST and HUT
spectra as for the IUE spectra.

The accurate measurement of the continuum flux is important.
Usually, the local continuum fitting with a power-law or a straight
line can produce reasonable measurement if the fitting is made with
caution. Here we use a straight line to fit the local continuum in
the selected continuum windows, namely 1260--1290\,{\AA},
1420--1460\,{\AA}, and 1805--1835\,{\AA}. The results are
satisfactory. In fitting the continuum, we find that the C~{\sc ii}
$\lambda$1334\,{\AA} line has relatively strong effects on the
continuum flux measurement at 1350\,{\AA}. Therefore, we adopt the
continuum flux at 1440\,{\AA} for our study of the line-continuum
relation of NGC 4151 because the lines around 1440\,{\AA}, like
O~{\sc iv}] $\lambda$1402\,{\AA} and N~{\sc iv}]
$\lambda$1486\,{\AA}, are relatively weak. The continuum flux  at
1440\,{\AA} was taken as the weighted mean over the wavelength range
1420-1460\,{\AA}, and the flux uncertainty is estimated from its
standard deviation.  No correction is made for the extinction, as it
is negligible for NGC 4151 (Kriss et al. 1992).

In the IUE spectra of NGC 4151, the \civ~$\lambda\lambda$1548.2,
1550.8 lines are blended with a few variable components covering a
wavelength range from 1480\,{\AA} to 1700\,{\AA} (see Clavel et al.
1987 for the details). To reduce the contamination from other lines,
we fit the continuum-subtracted spectra within the
wavelength range from 1450\,{\AA} to 1720\,{\AA} simultaneously with ten individual
Gaussian components. Five lines like N~{\sc iv}]
$\lambda$1486\,{\AA}, L1 $\lambda$1518\,{\AA}, L$'$2
$\lambda$1576\,{\AA}, L2 $\lambda$1594\,{\AA}, O~{\sc iii}]
$\lambda$1663\,{\AA} are fitted with a single Gaussian component.
The \civ~line is fitted with three Gaussian components: a broad
emission component, a narrow emission component, and a narrow
absorption component. The H~{\sc ii} $\lambda$1640\,{\AA} emission
line is fitted with two Gaussian components including a narrow one
and a broad one. Two narrow and variable satellite emission lines
(L1 and L2) were first noticed by Ulrich et al. (1985) when NGC 4151
was at a low continuum state. Using the high-resolution HST/STIS
spectra of NGC 4151, Crenshaw et al. (2000) indicate that these
satellite emission lines (L1, L$'$2, and L2) shown in the IUE
spectra are not in fact separate lines, but rather the residual
emission that appears isolated due to the low-ionization absorption
lines (primarily Si~{\sc ii}\,$\lambda$1526.7\,{\AA}, Si~{\sc
ii}$^*\,\lambda$1533.4\,{\AA}, and Fe~{\sc ii} multiples) overlaying
the large part of wings of \civ. Assuming them to be the satellite
emission lines can slightly underestimate the total emission line
flux of \civ~in the low-continuum case. The effects of such a
simplification on our result will be addressed in Sect.~3.

The nonlinear least-square Levenberg-Marquart minimization method is
used in our spectral fitting. Unlike Clavel et al. (1987)
who fixed the line widths of some Gaussian components, we
allow the fitting parameters for all components, such as the line
central wavelength, velocity dispersion, and the area under a line,
to vary at certain ranges. The uncertainties of these parameters are
obtained using a method similar to the one adopted in Clavel et al.
(1991). Three examples of our multi-component fitting of the spectra
taken at the low, moderate, and high flux states with IUE, HST, \&
HUT, respectively are shown in Fig.~1.  The fittings are satisfactory
in general. However, sometimes the red peak around 1540--1560\,{\AA}
is too sharp to be fitted perfectly, especially for the spectra
taken by HST. This leads to a slight underestimation of the total
\civ~emission line flux but will not affect our result
significantly.

\begin{figure}
\begin{center}
  \includegraphics[angle=0,height=10cm,width=8.6cm]{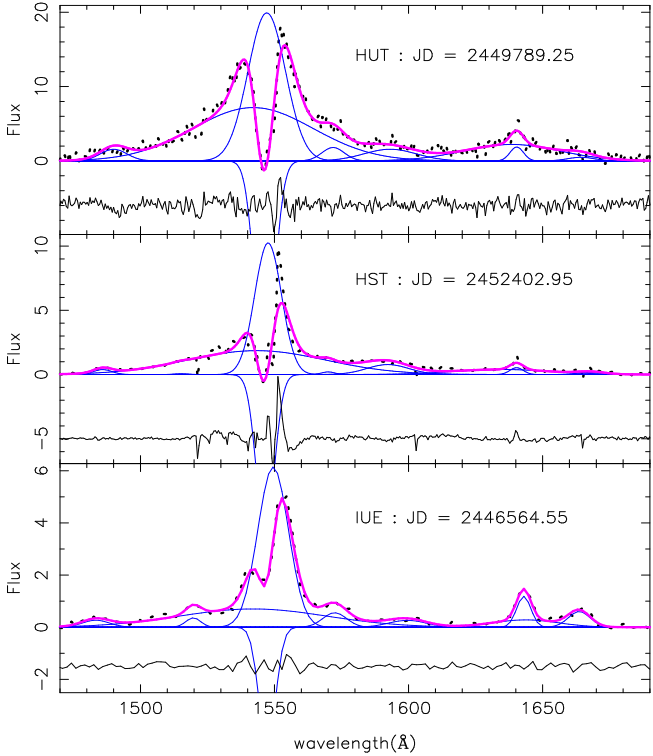}
  \caption{Sample spectra of NGC 4151 from IUE (lower panel), HST
(middle panel), \& HUT (upper panel) observations and fitting
results around the 1450--1720\,{\AA} region. The three spectra (from
bottom to top) are for the low, moderate, and high flux states,
respectively.  The dotted line represents the observed spectrum
after the continuum subtraction. The thin solid lines represent the
best-fit spectral components (see text for details). The thick solid
line denotes the sum of these components.  The residual of the fit
is also shown in the lower part of each panel.  Flux is in units of
$\rm 10^{-13}~erg~s^{-1}~cm^{-2}\AA^{-1}$.}
  \label{Fig.1}
\end{center}
\end{figure}

\begin{figure}
\begin{center}
  \includegraphics[angle=0,height=10cm,width=9.4cm]{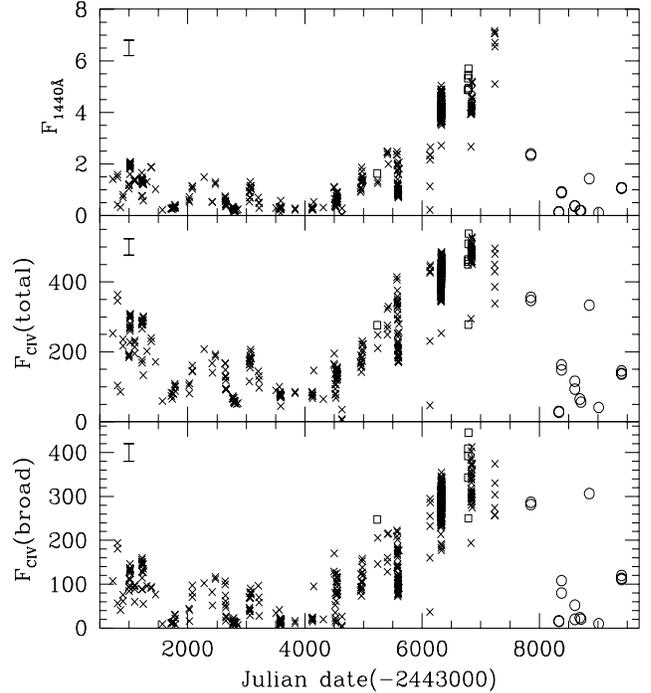}
   \caption{Light curves of the continuum flux at 1440\,{\AA}, the total
\civ~flux and the broad emission line component of NGC 4151 in
1978--2002. The crosses, squares, and circles represent the IUE,
HUT, and HST data, respectively. The average uncertainty was shown
in the upper-left part of each panel.  The continuum flux is in
units of $\rm 10^{-13}~erg~s^{-1}~cm^{-2}{\AA}^{-1}$ and the
\civ~line is in units of $\rm 10^{-13}~erg~s^{-1}~cm^{-2}$.}
  \label{Fig.2}
\end{center}
\end{figure}

In Fig. 2, we show the light curves of the continuum flux at
1440\,{\AA}, the total \civ~line flux (the sum of the modeled broad
emission, narrow emission, and absorption components), and the broad
emission-line component flux from the IUE, HUT, \& HST observations
in 1978--2002. A huge broad peak in 1991--1998 and several sub-peaks
are clearly shown in all three light curves. The overall response of
the \civ~emission line with the continuum variation is clear in
these long-term light curves. This also confirms that the
emission-line response with the continuum cannot be revealed by the
short duration observations (Crenshaw et al. 1996) but can be
clearly detected by the long-term spectral monitorings.


\section{The intrinsic Baldwin effect of NGC 4151}
Using the measured flux data for both \civ~emission line and UV
continuum, we can study the slope variation in the \civ~Baldwin
effect.

\begin{figure}
\begin{center}
  \includegraphics[angle=0,height=9cm,width=9.1cm]{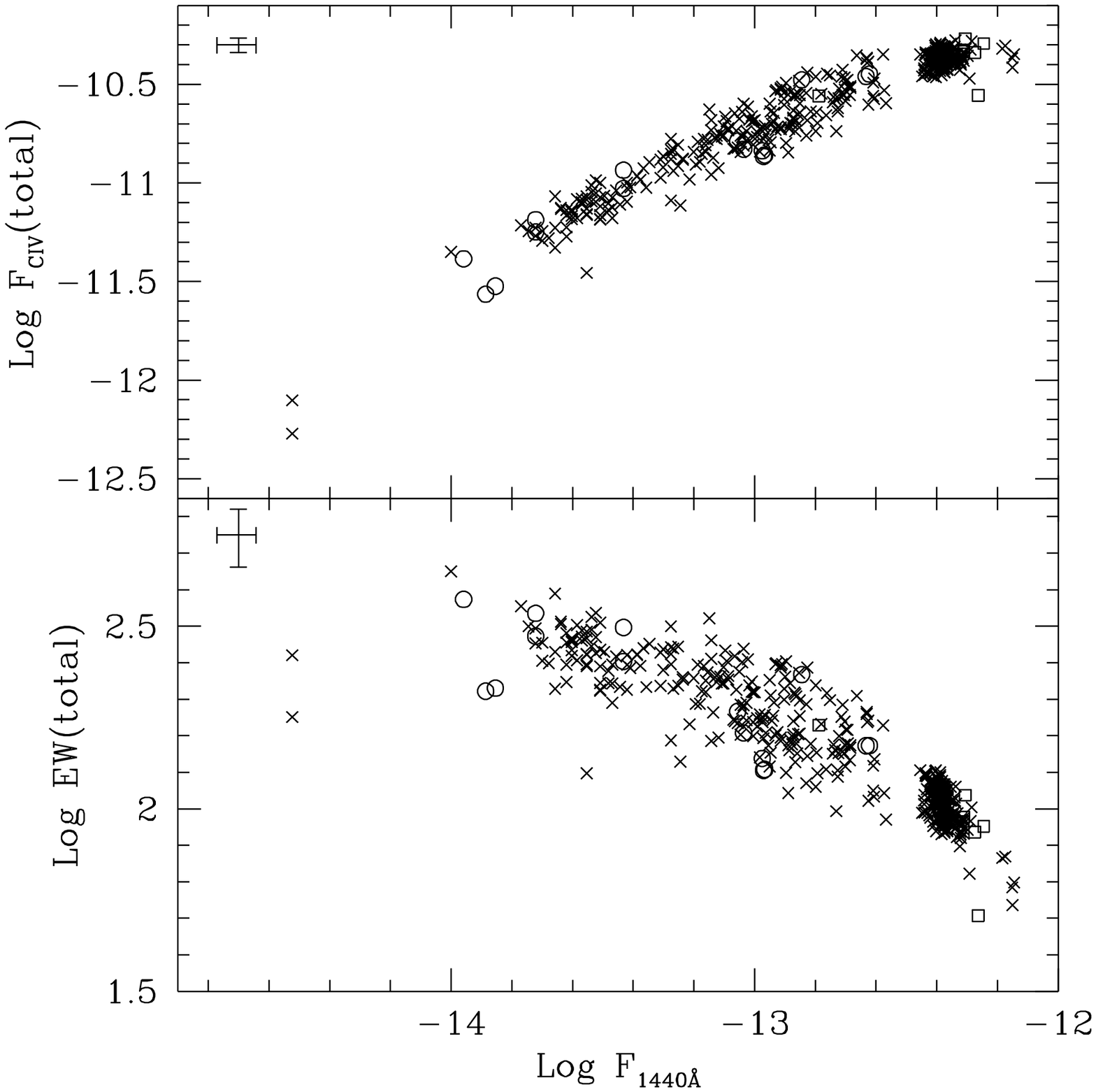}
  \caption{Correlations of the total \civ~line flux and the equivalent
width (EW) with the continuum flux at 1440\,{\AA} for NGC 4151.
Symbols have the same meanings as in Fig. 2. The average uncertainty
is shown in the upper-left part of each panel. The continuum flux,
line flux, and EW are in units of $\rm
erg~s^{-1}~cm^{-2}{\AA}^{-1}$, $\rm erg~s^{-1}~cm^{-2}$, and {\AA},
respectively.  The curvature of the Baldwin relationship is clearly
shown in both panels. The results from HUT \& HST are fully
consistent with those from IUE.}
  \label{Fig.3}
\end{center}
\end{figure}

\subsection{The non-constant slope}

The Baldwin effect represents the dependence of the equivalent
width (EW) of \civ~emission line on the UV continuum flux. Figure 3
shows the total \civ~emission-line flux and the line EW against the
UV continuum flux at 1440\,{\AA} for NGC 4151. The data from HUT \&
HST are fully consistent with those from IUE, clearly suggesting a
non-constant slope in the log-log plots.  We notice that the trend
of the curvature in the upper panel of Fig. 3 is similar to that
found for the intrinsic H$\beta$ Baldwin effect of NGC 5548 (see
Fig. 4 of Goad et al. 2004) and Fairall 9 (see Fig.~2 of Wamsteker \& Colina 1986),
 which also show the steepening of the
slope at the lower flux state and the flattening of the slope at the
higher flux state. In the  EW-continuum plot (the lower panel of
Fig. 3), the flattening of the slope in the low continuum state
 is also consistent with the trend noticed by Osmer \& Shields
(1999) for Fairall 9.  This suggests that
the non-constant slope of the Baldwin relationship may be usual for
AGNs.

To demonstrate the non-constant slope of the Baldwin effect for NGC
4151 more clearly, we divide the observations into 4 epochs. Epoch 1
(JD2443722--JD2445654) covers the IUE observations in 1978--1983,
which shows at least two sub-peaks in the light curve. Epoch 2
(JD2445766--JD2447631) covers the IUE observations in 1984--1989,
which represents the lowest flux state in all the UV observations.
Epoch 3 (JD2447948--JD2450243) covers the IUE observations in
1990--1996 and the HUT observations in 1990 and 1995. The number of
spectra in this epoch is significantly larger than others since it
includes 203 spectra obtained in a one-month intensive IUE campaign
in 1993, which covers the highest flux state of NGC 4151. Epoch 4
covers the HST observations in 1998--2002. Although only 15 HST/STIS
spectra are available, we can still clearly see that both the
continuum flux and \civ~emission line flux fade away in this epoch
from the major peak. The numbers of spectra, the average UV
continuum flux, and the estimated slope of the Baldwin effect in
these 4 epochs are listed in Table 1. The variation in the slope
$\alpha$ with the UV continuum flux is shown clearly in Fig. 4. The
slope $\alpha$ varies from 0.58 in epoch 3 (the highest flux case)
to 0.83 in epoch 2 (the lowest flux case).  The result from the HST
observations ($\alpha=0.78$ in epoch 4) confirms the trend obtained
from the IUE and HUT observations in the other 3 epochs. A smaller
slope $\beta$ value ($-$0.17) in epoch 3 is consistent with the
almost constant EW value at the lower flux state noticed for Fairall
9 by Wamsteker \& Colina (1986) and Osmer \& Shields (1999).
The slope $\alpha$ values for NGC 4151
are also well within the range for the broad H$_\beta$ line of NGC
5548 found by Goad et al. (2004), who indicated that the slope
$\alpha$ varies from 0.4 to 1 in the 13-year observations of NGC
5548.

\begin{table*}
\caption{The average continuum flux and slope of the Baldwin effect
in four different observation epochs.  The standard deviations of
these two values are also given. The flux is in units of $\rm
10^{-13}~erg~s^{-1}~cm^{-2}\AA^{-1}$.  } \label{table1}
$$
\begin{array}{cccccccc}
\hline
\noalign{\smallskip}
\rm{Epoch} & \rm{JD~~Time} & \rm{No.~of~Spectra} &F_{1440\AA} & \sigma(F_{1440\AA}) & \alpha &
\beta & \sigma
\\
\noalign{\smallskip}
\hline
1& \rm{JD2443722-JD2445654} & \rm{75(IUE)} & 1.077 & 0.574 & 0.721 & -0.279 & 0.025\\
2& \rm{JD2445766-JD2447631} & \rm{85(IUE)} & 0.490 & 0.307 & 0.833 & -0.167 & 0.059\\
3& \rm{JD2447948-JD2450243} & \rm{308(IUE)+7(HUT)}& 3.607 & 1.294 & 0.581 & -0.419 & 0.025\\
4 & \rm{JD2450855-JD2452403} & \rm{15(HST)}& 0.844 & 0.754 & 0.777 & -0.223 & 0.056\\
\noalign{\smallskip}
\hline
\end{array}
$$
\end{table*}

\begin{figure}
\begin{center}
  \includegraphics[angle=0,height=8.5cm,width=8.5cm]{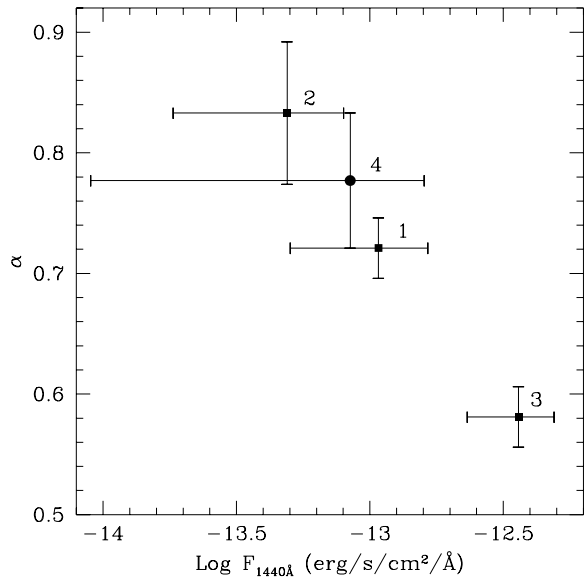}
  \caption{Variation of the slope $\alpha$ with the continuum flux for
NGC 4151 in 4 observation epochs. The epoch number is shown for each
point.  The epoch 4 is for the HST observations in 1998--2002. See
Table 1 for details.  }
  \label{Fig.4}
\end{center}
\end{figure}

\begin{figure}
\begin{center}
  \includegraphics[angle=0,height=10cm,width=9.2cm]{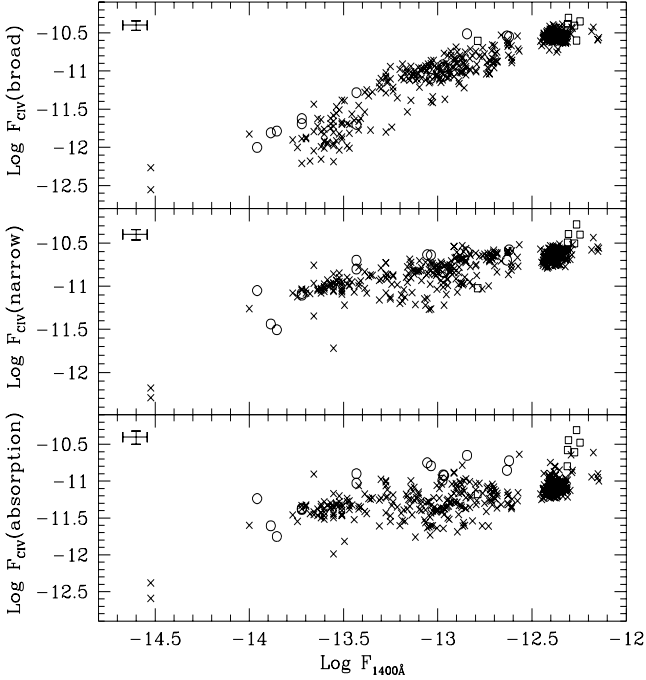}
  \caption{Variations in the broad, narrow, and absorption components
of \civ line with the UV continuum flux at 1440\,{\AA} for NGC 4151.
Symbols have the same meanings as in Fig. 2. The average errors are
shown in the upper-left part of each panel. The continuum
flux and line flux are in units of $\rm erg~s^{-1}~cm^{-2}\AA^{-1}$ and $\rm
erg~s^{-1}~cm^{-2}$ respectively.  }
  \label{Fig.5}
\end{center}
\end{figure}

\subsection{The effects of absorption features}

The non-constant slope of the intrinsic Baldwin effect has been
found so far for NGC 4151 (this paper), Fairall 9 (Wamsteker
\& Colina 1986; Osmer \& Shields
1999), and NGC 5548 (Goad et al. 2004).
However, unlike Fairall 9
and NGC 5548, which show no absorptions and weak ones in their UV
spectra, respectively (Crenshaw et al. 1999), NGC 4151 shows much
more remarkable absorption features, especially in the \civ~line
profile (Boksenberg et al. 1978; see also Fig. 1).  These absorption
components also vary with the UV continuum.  From Fig. 1 we see that
the central absorption is stronger when the continuum flux is
higher.  This absorption feature has been found to actually combine
various absorption lines with different ionization potentials (Kriss
et al. 1992; Weymann et al. 1997; Crenshaw et al. 2000; Kraemer et
al. 2005).  Considering the absorption feature, a natural question
then arises: is the non-constant slope of the \civ~Baldwin effect in
NGC 4151 due to the variation in absorption at different continuum
flux levels?  To answer this question, we investigated the
variations in the broad emission, narrow emission, and absorption
components of \civ~line. Figure 5 shows the variations in these
components with the UV continuum flux.  We can see that the flux
values of all these three components decrease when the continuum
flux decreases. The variation in the absorption component is very
similar to that of the narrow emission-line component. Moreover, we
noticed that in the higher continuum flux state, the broad-line
component always dominates the total \civ~flux. In the lower
continuum flux state, however, the narrow emission-line component
becomes dominant. Both the broad-line component and the central
absorption components become relatively weak.
 If we consider the Baldwin
effect of the broad emission-line component alone, we also can
obtain a non-constant slope with $\alpha$ varying from 0.74$\pm$0.02
in the highest flux epoch to 1.33$\pm$0.12 in the lowest flux epoch,
which clearly indicates that the non-constant slope may not be
dominated by the absorptions.

As mentioned in Sect.~2, besides the remarkable central absorption
component, there are also many subtle absorption features in the
wings of \civ~profile of NGC 4151, as revealed by the
high-resolution HST spectra (Weymann et al. 1997; Crenshaw et al.
2000; Kraemer et al. 2001). These complex features make it more
difficult to extract the information of the unabsorbed \civ~flux. In
our spectral fitting, we only use a Gaussian component to represent
the central absorption of \civ~and treat the absorption-induced
`emission' features in the wings of \civ~as satellite emission lines
(L1, L$'$2, and L2). As these features are in fact caused by the
low-ionization absorption lines overlaying the large part of wings
of \civ, our simplification would underestimate the total flux of
\civ. As indicated by Crenshaw et al. (2000), these low-ionization
absorption features are weak in the high-flux state and become
prominent when the UV continuum drops to a low state. Therefore, the
simplification of our spectral fitting is fine when NGC 4151 is
bright (e.g. the continuum flux larger than $\rm
10^{-13}\,erg~s^{-1}~cm^{-2}\AA^{-1}$,
 as in epochs 1 and 3). Even from the result obtained in these two epochs
when NGC 4151 is relatively bright, the slope variation of the
Baldwin effect is still evident (see Fig. 4). Using the {\it
HST/STIS} spectra, we also investigate the difference in the
spectral fitting with our model and the model consisting of three
Gaussian components to represent three low-ionization absorption
lines at around 1526\,\AA, 1533\,\AA, and 1577\,{\AA}, respectively.
The difference is about 2\% when the UV continuum flux is larger
($F_{1440\AA}=2.4\times \rm 10^{-13}\,erg~s^{-1}~cm^{-2} \AA^{-1}$
on JD2450855) and about 13\% when the UV continuum flux is smaller
($F_{1440\AA}=1.4\times \rm 10^{-14}\,erg~s^{-1}~cm^{-2} \AA^{-1}$
on JD2451334). Although the equivalent widths of these
low-ionization absorption lines become greater when the source is
fainter, we see that the simplification of our spectral fitting does
not lead to significant underestimation of the total \civ~flux even
in the faint state of NGC 4151. Together with the similar trend in
the Baldwin effect found for Fairall 9, which does not show
absorptions in \civ, and for NGC 5548, which does show some narrow
absorption features in \civ~but not in H$_\beta$, we believe that
the non-constant slope of the \civ~Baldwin effect in NGC 4151 should
not be driven mainly by the absorption effect. However, more
quantitative studies of the effects of complex absorption features
on the Baldwin effect of NGC 4151 are still needed.


\section{Conclusion and discussion}

With the long-term UV spectral data  of IUE, HUT, \& HST, we found
the non-constant slope in the intrinsic \civ~Baldwin effect of
Seyfert 1 galaxy NGC 4151. The trend of the slope change with the UV
continuum flux variation is similar to those found for two other
Seyfert 1s, Fairall 9 and NGC 5548, suggesting that the non-constant
slope may be usual for AGNs. The physical origin of such a
non-constant slope is probably related to  the different response of
the broad-line emission to the continuum variations at different
luminosity levels (Korista et al. 1998).

From the theoretical point of view, an accretion disk at different
luminosity levels (corresponding to different accretion rates)
should have different accretion modes. Probably the AGNs accrete in
a radiatively inefficient accretion flow (see Narayan, Mahadevan \&
Quataert 1998 for a review) at a lower accretion rate and in a
standard optically thick disk (Shakura \& Sunyaev 1973) or a slim
disk (Abramowicz et al. 1988) at a higher accretion rate. Different
accretion modes radiate differently, producing a relatively hard
spectrum at a lower accretion rate and a soft spectrum at a higher
accretion rate. Because the ionizing continuum of AGNs mainly comes
from the radiation of the accretion disk, the difference in the
disk-emitting spectrum may lead to the different response of the
emission line to the ionizing continuum through the photoionization
process. For NGC 4151, the lowest and highest continuum fluxes at
1440\,{\AA} in the 1978--2002 observational period are about $\rm
3\times10^{-15}\,erg~s^{-1}cm^{-2}\AA^{-1}$ and $\rm
6\times10^{-13}\,erg~s^{-1}cm^{-2}\AA^{-1}$, which correspond to
luminosities at 1440\,{\AA} of \rm $\rm 10^{41}\,erg~s^{-1}$ and
$\rm 2\times 10^{43}\,erg~s^{-1}$, respectively. If we adopt the
black-hole mass for NGC 4151 as $\rm 1.33\times 10^7M_\odot$
(Peterson et al. 2004) and assume $f_\nu\propto \nu^{-0.5}$ in the
UV/optical band and
 $L_{bol}\simeq 9L_{5100\,\AA}$ for deriving the bolometric
luminosity (Kaspi et al. 2000), we can estimate the Eddington ratio
($L_{bol}/L_{Edd}$, usually taken as a measure of the dimensionless
accretion rate) in the lowest and highest flux states of NGC 4151 as
0.001 and 0.2, respectively. Clearly, the dimensionless accretion
rate of NGC 4151 varies more than two orders of magnitude. Because
the critical dimensionless accretion rate between a radiatively
efficient accretion flow and a radiatively inefficient one is about
0.01 (Narayan et al. 1998), most probably NGC 4151 accretes in a
radiatively inefficient accretion flow at the lower flux state but
in a radiatively efficient one at the higher flux state. Such a
change of accretion modes can not only produce different spectral
energy distributions in the ionizing continuum, but can also produce
different non-linear responses of the \civ~emission line and  then
lead to the non-constant slope of the Baldwin effect in the
different luminosity levels.

Previous studies have revealed the importance of the light-travel
time effect on the Baldwin effect (Krolik et al. 1991; Pogge \&
Peterson 1992; Peterson et al. 2002). Correction of such an effect
can substantially reduce the scatters of the intrinsic Baldwin
effect. However, the time lag for NGC 4151 between the \civ~emission
line and the UV continuum flux is not well-determined. Clavel et al.
(1990) obtained a delay of $3.2\pm 3$ days using the two-month data
with a mean interval of 3.4 days of the IUE campaign in 1988--1989
when NGC 4151 was relatively faint.  Crenshaw et al. (1996) failed
to detect any correlation between the \civ~line and UV continuum
flux from a short but intensive IUE campaign in December 1993 when
NGC 4151 was relatively bright. We also used the whole IUE data set
in 1978--1996 and estimated the time lag between the total \civ~line
flux and the continuum flux at 1440\,{\AA} with an interpolation
cross-correlation function (ICCF) method developed by Gaskell \&
Sparke (1986), Gaskell \& Peterson (1987), and White \& Peterson
(1994). We found that the time lag is about 1.9 days. We tried to
use such a time lag to correct the light-travel time effect for NGC
4151 and found that our result does not change. This is mainly
because the mean interval of the whole UV data set in 1978--2002 is
about 15 days, which is significantly longer than the estimated time
lag.

In conclusion, we find the non-constant slope of the intrinsic
Baldwin effect in NGC 4151 and suggest that such a non-constant
slope may not be unusual for AGNs. Obviously, more studies on more
AGNs are needed to confirm our result. This will require intensive,
long-term, and high-resolution UV spectral monitoring with the
current and future space UV observatories on some strongly variable
AGNs. We expect that these future studies will help us to extract
more accurate emission-line and continuum fluxes and establish a
more reliable line-continuum relation for AGNs.


\begin{acknowledgements}

We are grateful to Prof. Brad Peterson for kindly providing an ICCF
program to calculate the time lag, to the anonymous referee for
his/her constructive comments which help to improve the paper
significantly, and to Mr. Lei Qian and Mr. Bingxiao Xu for many
helpful discussions. The authors are supported by the National
Natural Science Foundation of China (Grants No. 10473001, No.
10525313, and No. 10521001), the RFDP Grant (No. 20050001026), and
the Key Grant Project of Chinese Ministry of Education (No. 305001).

\end{acknowledgements}


\begin{thebibliography}{}

\bibitem[1988]{Abramowicz}Abramowicz, M. A., Czerny, B., Lasota, J.
P., \& Szuszkiewicz, E. 1988, ApJ, 332, 646
\bibitem[2004]{}Bachev, R., Marziani, P., Sulentic, J. W., Zamanov,
 R., Calvani, M., \& Dultzin-Hacyan, D. 2004, ApJ, 617, 171
\bibitem[]{}Baldwin, J. A. 1977, ApJ, 214, 679
\bibitem[]{}Baldwin, J. A., Wampler, E. J., \& Gaskell, C. M. 1989, ApJ,
338, 630
\bibitem[]{}Baskin, A., \& Laor, A. 2004, MNRAS, 350, 31
\bibitem[]{}Boksenberg, A., Snijders, M. A. J., Wilson, R., et
al. 1978, Nature, 275, 404
\bibitem[Boroson \& Green (1992)]{bg92} Boroson, T. A., \& Green,
R. F. 1992, ApJS, 80, 109
\bibitem[1987]{Clavel87}Clavel, J., Altamore, A., Boksenberg, A., et
al. 1987, ApJ, 321, 251
\bibitem[1990]{clavel90}Clavel, J., Boksenberg, A., Bromage, G. E., et
al. 1990, MNRAS, 246, 668
\bibitem[1991]{Clavel91}Clavel, J., Reichert, G. A., Alloin, D., et
al. 1991, ApJ, 366, 64
\bibitem[]{}Crenshaw, D. M., Kraemer, S. B., Boggess, A., Maran, S. P., Mushotzky, R. F., \& Wu,
C.-C. 1999, ApJ, 516, 750
\bibitem[2000]{Crenshaw00}Crenshaw, D. M., Kraemer, S. B., Hutchings,
J. B., et al. 2000, ApJ, 545, 27
\bibitem[1996]{Crenshaw96} Crenshaw, D. M., Rodriguez-Pascual, P. M.,
Penton, S. V., et al. 1996, ApJ, 470, 322
\bibitem[1996]{Edelson96}Edelson, R. A., Alexander, T., Crenshaw, D. M.,
et al. 1996, ApJ, 470, 364
\bibitem[]{}Gaskell, C. M. \& Peterson, B. M. 1987, ApJS, 65, 1
\bibitem[]{}Gaskell, C. M. \& Sparke, L. S. 1986, ApJ, 305, 175
\bibitem[]{}Gilbert, K. M., \& Peterson, B. M. 2003, ApJ,
587, 123
\bibitem[]{}Goad, M. R., Korista, K. T., \& Knigge, C. 2004, MNRAS, 352,
277
\bibitem[]{}Jones, B. J. T., \&  Jones, J. E. 1980, MNRAS, 193, 537
\bibitem[Kaspi et al. (2000)]{k00} Kaspi, S., Smith, P. S.,
Netzer, H.,  Maoz, D., Jannuzi, B. T., \& Giveon, U. 2000, ApJ, 533,
631
\bibitem[]{}Kinney, A. L., Huggins, P. J., Glassgold, A. E., \& Bregman,
J. N. 1987, ApJ, 314, 145
\bibitem[]{}Kinney, A. L., Rivolo, A. R., \& Koratkar, A. P. 1990, ApJ, 357, 338
\bibitem[]{}Korista, K., Baldwin, J., \& Ferland, G. 1998, ApJ, 507, 24
\bibitem[]{}Kraemer, S. B., Crenshaw, D. M., Hutchings, J. B., et
al. 2001, ApJ, 551, 671
\bibitem[]{}Kraemer, S. B., George, I. M., Crenshaw, D. M., et
al. 2005, ApJ, 633, 693
\bibitem[1992]{Kriss}Kriss, G. A., Davidsen, A. F., Blair, W. P.,
et al. 1992, ApJ, 392, 485
\bibitem[]{}Kriss, G. A., Davidsen, A. F., Zheng, W., Kruk, J. W., \& Espey, B. R. 1995,
ApJ, 454, 7
\bibitem[]{}Krolik, J. H., Horne, K., Kallman, T. R., Malkan, M. A., Edelson, R. A., \& Kriss, G. A. 1991,
ApJ, 371, 541
\bibitem[]{}Murdoch, H. S. 1983, MNRAS, 202, 987
\bibitem[Narayan, Mahadevan \& Quataert (1998)]{n98}Narayan, R.,
Mahadevan, R., \& Quataert, E., 1998, In Theory of Black Hole
Accretion Disks, eds M. A. Abramowicz, G. Bjornsson, \& J. Pringle,
Cambridge University Press, p.148
\bibitem[]{}Netzer, H., Laor, A., \& Gondhalekar, P. M. 1992, MNRAS, 254, 15
\bibitem[]{}Osmer, P. S., \& Shields, J. C. 1999, ASPC, 162, 235
\bibitem[]{}Peterson, B. M., Berlind, P., Bertram, R., et al. 2002,
ApJ, 581, 197
\bibitem[]{}Peterson, B. M., Ferrarese, L., Gilbert, K. M., et
al. 2004, ApJ, 613, 682
\bibitem[]{}Pogge, R. W., \& Peterson, B. M. 1992, AJ, 103, 108
\bibitem[Shakura \& Sunyaev (1973)]{ss73} Shakura, N. I., \& Sunyaev,
R. A. 1973, A\&A, 24, 337
\bibitem[]{}Shang, Z.-H., Wills, B. J., Robinson, E. L., et al. 2003,
ApJ, 586, 52
\bibitem[1996]{Ulrich96}Ulrich, M.-H. 1996, MNRAS, 281, 907
\bibitem[]{}Ulrich, M.-H. 2000, A\&ARv, 10, 135
\bibitem[]{}Ulrich, M. H., Altamore, A., Perola, G. C., et al. 1985, Nature, 313, 747
\bibitem[]{}Ulrich, M.-H., Boksenberg, A., Penston, M. V., et
al. 1991, ApJ, 382, 483
\bibitem[]{}Wamsteker, W., \& Colina, L. 1986, ApJ, 311, 617
\bibitem[]{}Warner, C., Hamann, F., \& Dietrich, M. 2004, ApJ, 608, 136
\bibitem[]{}Weymann, R. J., Morris, S. L., Gray, M. E., \& Hutchings, J. B. 1997,
ApJ, 483, 717
\bibitem[]{}White, R. J., \& Peterson, B. M. 1994, PASP, 106, 879
\bibitem[]{}Wu, C.-C., Boggess, A., \& Gull, T. R. 1983, ApJ, 266, 28
\bibitem[]{}Zheng, W., Kriss, G. A., Telfer, R. C., Grimes, J. P., \& Davidsen, A. F. 1997, ApJ,
475, 469

\end{thebibliography}
\end{document}